\documentstyle[twoside,fleqn,espcrc2]{article}
\input epsf
\epsfverbosetrue

\newcommand{\AmS}{{\protect\the\textfont2
  A\kern-.1667em\lower.5ex\hbox{M}\kern-.125emS}}

\title{Flux-tube Structure, 
Sum Rules and Beta-functions in SU(2) \thanks{Presented by A.M.~Green, {\tt green@phcu.helsinki.fi}}}

\author{A.M.~Green, P.~Pennanen\address{Department of Physics and The 
Helsinki Institute of Physics, University of Helsinki, Finland }
        and 
        C.~Michael\address{Theoretical Physics Division, 
Dept. of Math. Sciences, University of Liverpool, Liverpool, UK}
\thanks{\tt petrus@hip.fi, cmi@liv.ac.uk}}
       
\begin{document}

\begin{abstract}
Action and energy flux-tube profiles are computed, in SU(2) with
$\beta=2.4,2.5$, for two quarks up to 1 fm apart and for which the colour 
fields are in their ground state ($A_{1g}$) and the first ($E_{u}$) and 
higher ($A'_{1g}$) excited gluonic states. When these profiles are integrated
over all space, a scaling comparison is made between the $\beta=2.4$ and
2.5 data. Using sum rules, these integrated forms also permit an estimate to be
made of generalised $\beta$-functions giving $b(2.4)=-0.312(15)$,
$b(2.5)=-0.323(9)$, $f(2.4)=0.65(1)$ and $f(2.5)=0.68(1)$. When the profiles
are integrated only over planes transverse to the interquark line and
assuming underlying string features, scaling comparisons are again made 
near the centres of the interquark line for the largest interquark distances.
For the $A'_{1g}$ case, some of the profiles exhibit a 
\lq{}dip-like\rq $\;$structure characteristic of the Isgur-Paton model.
\end{abstract}

\maketitle

Energy and action profiles for two quarks a distance $R$ apart are calculated
by measuring the correlation $<WP^{\mu \nu}>$ between the Wilson loop $W(R)$ and
different orientations of the plaquette $P^{\mu \nu}$.
One of the new features compared with earlier such studies is that here the
colour fields are not only in the ground state $(A_{1g})$ but also in the
excited states $(E_u)$ and $(A'_{1g})$ -- more details being given in refs.
\cite{gmp,pgm}.

The colour fields are measured using a plaquette, whose physical size
changes with $\beta$. As only observables with the
same physical size at different values of the coupling have a continuum
limit, there will be no naive scaling but we are able
to make use of the sum-rules presented below (after subtracting
divergent self-energy contributions) to control the normalisation of the
three-dimensional sums of the fields. More microscopic observables, such as 
planar sums or transverse
profiles, do not have a well defined continuum limit, so our comparisons
for these at the two values of coupling should be taken as exploratory.
                             
Figs. 1, 2 and 3 show for the three gluonic states the action $(S)$, 
Longitudinal and Transverse energies $(E_{L,T})$ profiles.  Here the 
$\beta=2.4$  data has been scaled by the factor $2.4/(2.5\rho^4)$, where
$\rho=a(2.4)/a(2.5)=1.410(13)$. It is seen that, except for $S(A_{1g})$ and
 $E_L(A_{1g})$, \lq{}scaling\rq $\;$is not very evident. The effect of discretization 
means that any underlying profile structure
tends to get \lq{}smoothed out\rq $\;$as $\beta$ decreases, since the lattice spacing is
larger. This is seen in the figures as sharper peaks for the $\beta=2.5$ data.

Figs. 4, 5 and 6 show the profiles, integrated over transverse planes  -- 
again for the three gluonic states. Several features are apparent:

i) The peaks at $R_L\approx 6$ are the self-energies at the quark
positions. They diverge in a manner suggested by
leading order perturbation theory.

ii) For $0\leq R_L\leq 4$ clear flux-tubes emerge each with a constant radius.

iii) \lq{}Scaling\rq $\;$between the $\beta =2.4$ and 2.5 data is now clearer than
in Figs. 1 and 2.

iv) The $E_T$ sum increases when going from the $A_{1g}$ to $E_u$ to the
 $A'_{1g}$ gluonic states -- a feature expected from a vibrating string model.

When the transverse sums are themselves integrated over $R_L$ to give
the total energy or action, a comparison can be made with the sum rules:
 \begin{eqnarray} {-1 \over b} \left( V+R {\partial V \over \partial R}
\right) + S_0 & = & \sum S 
 \label{TASU} \\
{1 \over 4 \beta f} \left( V+R {\partial V \over \partial R} \right) +
E_0 & = & \sum E_L
\label{TELSU} \\
{1 \over 4 \beta f} \left( V-R {\partial V \over \partial R} \right) +
E_0 & = & \sum E_T.
\label{TEPSU} 
\end{eqnarray}
Here $V(R)$ is the interquark potential and $b,f$ are generalised 
$\beta$-functions, which show
how the bare  couplings of the theory vary with the generalised lattice
spacings $a_\mu$ in four  directions. The main interest is to extract
$b,f$
by first calculating $V$ and the $\sum $'s on a lattice.
 Unfortunately, this
strategy is complicated by two features -- the unknown self-energies 
$(S_0,E_0)$  and the value of $\partial V/\partial R$.
Here three methods are attempted to estimate $b,f$ \cite{pgm}:  

\vskip 0.2 cm

\noindent\underline{Method 1}. Since $V$ is known numerically,                                            
$ V\pm R {\partial V \over \partial R}$ can be calculated and plotted
against, say $\sum S$. This is a linear plot and the slope gives $b$.
The extraction of $f$ is less clean and necessitates a simultaneous 
fit using both the $\sum E_L$ and $\sum E_T$ sum-rules. This arises
because the $E_{L,T}$ are {\em differences} between the electric and
magnetic fields -- unlike the action -- and this leads to a numerical
inaccuracy problem. The outcome yields our best estimates of $b,\,f$.

\vskip 0.2 cm

\noindent\underline{Method 2}. In the large $R$ limit $V(R)=-e/R+b_sR+V_0$, so
that   ${\partial V \over \partial R}$ is readily estimated.
Furthermore, the self-energies can be removed by using the sum-rules at
two different values of $R=R_1,R_2$ to give:
\begin{eqnarray}
 b & = & \frac{-2b_S(R_1-R_2)}{\sum S_{R_1} - \sum S_{R_2}}
\label{elrb}\\
 f(I) & = & \frac{b_S(R_1-R_2)}{2\beta [\sum (E_L)_{R_1} - \sum
(E_L)_{R_2}]} \\
 f(II) & = & \frac{-e(1/R_1-1/R_2)}{2\beta [\sum (E_T)_{R_1} - \sum
(E_T)_{R_2}]}.
\end{eqnarray}
Using directly the 3-D sums $\sum (S,E)_R$ gives results consistent with
Method 1 but having larger error bars. The further approximation of a constant 
longitudinal flux tube profile gives consistent results at the largest $R$'s, 
supporting a string-like picture. 

\vskip 0.2 cm

\noindent\underline{Method 3}. It is possible to combine the three sum-rules
using two values of $R$ in such a way as to completely eliminate both
${\partial V \over \partial R}$  and the self-energies.
This would seem ideal, since it involves quantities that can be measured directly.
However, in practice, there is a problem, since now even $b$ depends on
the energy differences, leading to
large uncertainties. 

\vskip 0.2 cm

Our best values for $b$ shown in the abstract can be compared 
with other recent estimates in Table 1, most importantly the finite T approach of
Ref. \cite{eng:95}. Based on the agreement of non-perturbative estimates we 
conclude that order $a^2$ effects in the
extraction of the $\beta$-function are small at the $\beta$-values
studied using the methods described. Thus a unique
$\beta$-function describes the deviations from
asymptotic scaling at these values of the coupling.

\begin{table}[htb]
\vspace{-0.8cm}
\begin{center}
\begin{tabular}{lcccc} 
$\beta$  & Ref. \cite{eng:95} & Ref. \cite{pen:96b} & 3-loop PT \\ \hline
2.4      & -0.3018  & -0.305(6  & -0.3893 \\
2.5      & -0.3115  & -0.312(2) & -0.3889 \\ 
\end{tabular}
\caption{Comparison between values of $b \equiv \partial \beta/\partial \ln a$
\label{tb}}
\end{center}
\end{table}

\vspace{-0.7cm}

A Viennese group has measured field distributions around a static source pair 
in dually transformed U(1) 
on a lattice \cite{zac:97}. Their results for the three-dimensional sums vs. 
quark separation show slopes and ordering of the transverse and 
longitudinal components of electric and magnetic fields similar to ours. 

At present theories \cite{IP,BBZ} only calculate the total energy
profile for the $A_{1g}$ state and for this they show some success.
However, no one has a profile model for $S$ or $E_{L,T}$ or for the excited
gluon fields. Such comparisons will be very demanding for any model
proposed.

\begin{figure}[p]
\vspace{-1.5cm}
\hspace{0cm}\epsfysize=140pt\epsfxsize=200pt\epsfbox{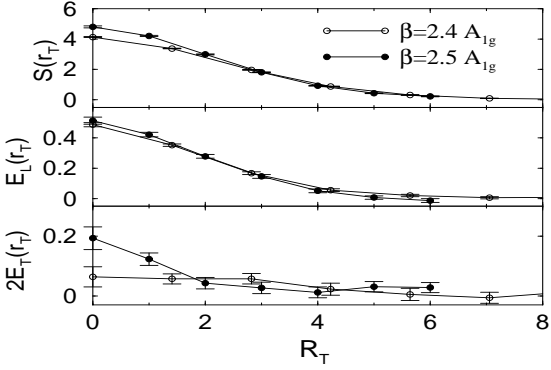}
\vspace{-0.8cm}
 \caption{ The colour flux contributions for $S$, $E_{L,T}$.
These are shown in units of $a(2.5)$ versus transverse
distance $R_T$ at the mid-point ($R_L=R/2$) for $\beta=2.4,\,2.5$ and
separation $R=8,12$ i.e. 0.95, 1.01 fm respectively.
The data are for the symmetric  ground state
(A$_{1g}$ representation), multiplied with a factor of $10^3$.}
 \label{frt}
\end{figure}
\begin{figure}[p]
\vspace{-1.2cm}
\hspace{0cm}\epsfysize=140pt\epsfxsize=200pt\epsfbox{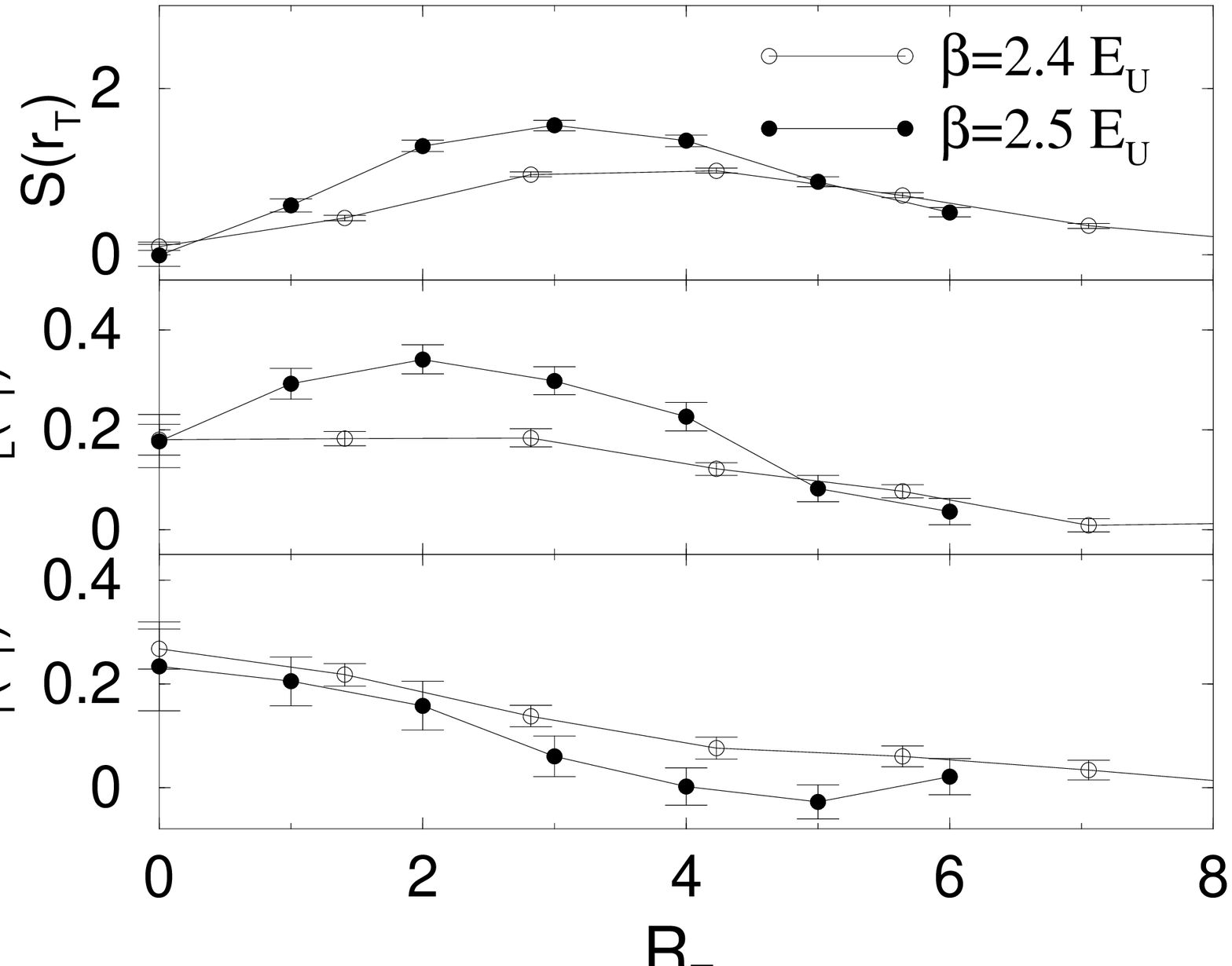}
\vspace{-0.8cm}
 \caption{ As Fig.~\protect\ref{frt} but for the first gluonic
excitation ($E_u$ representation).}
 \label{frte}
\end{figure}
\begin{figure}[p]
\vspace{-1.2cm}
\hspace{0cm}\epsfysize=140pt\epsfxsize=200pt\epsfbox{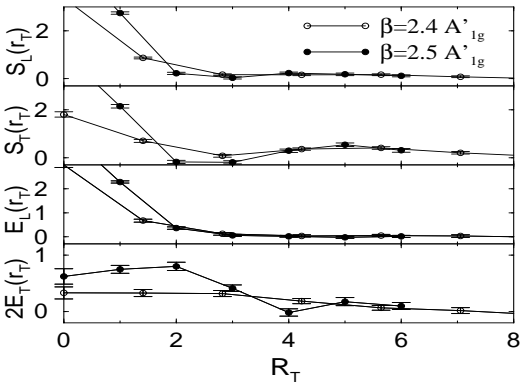}
\vspace{-0.8cm}
 \caption{ As Fig.~\protect\ref{frt} but for $R=4,6$ and the higher gluonic
excitation (first excited state of $A_{1g}$ representation).}
 \label{frtp}
\end{figure}

\begin{figure}[p]
\vspace{-2cm}
\hspace{0cm}\epsfysize=150pt\epsfxsize=220pt\epsfbox{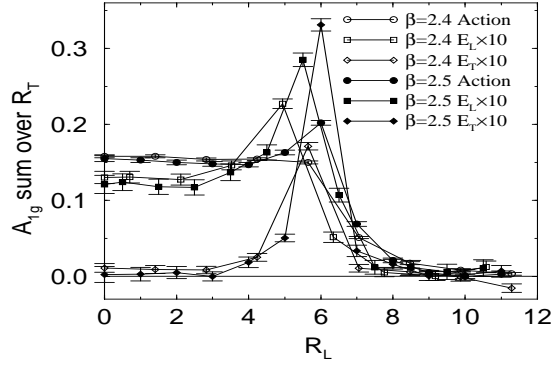}
\vspace{-1cm}
\caption{The dependence on longitudinal position ($R_L$) of the sum
over the transverse plane of the colour flux  contributions
for $S,E_{L,T}$.
Here $R_L$ is measured from the mid-point for the same case as in Fig. 1.}
\label{frl}
\end{figure}
 \begin{figure}[p]
\vspace{-2cm}
\hspace{0cm}\epsfysize=150pt\epsfxsize=220pt\epsfbox{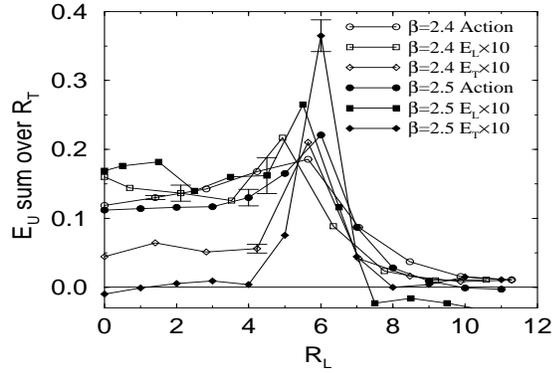}
\vspace{-1cm}
\caption{As Fig.~\protect\ref{frl} but for the first gluonic
excitation. For each data set one error bar is
shown.}
 \label{frle}
\end{figure}
 \begin{figure}[p]
\vspace{-2cm}
\hspace{0cm}\epsfysize=150pt\epsfxsize=220pt\epsfbox{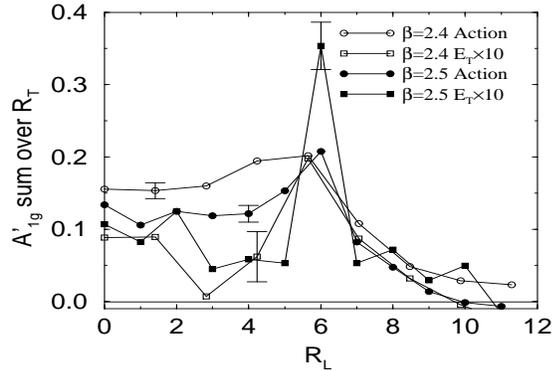}
\vspace{-1cm}
\caption{As Fig.~\protect\ref{frle} but for the higher gluonic
excitation. }
 \label{frlp}
\end{figure}

\end{document}